# Kilometer-precise (UII) Umbriel physical properties from the multichord stellar occultation on 2020 September 21

M. Assafin[1,2]★, S. Santos-Filho[1,2,3], B, E. Morgado[1,2,9], A. R. Gomes-Júnior[5,2], B. Sicardy[6], G. Margoti[7,2], G. Benedetti-Rossi[8], F. Braga-Ribas[7,2], T. Laidler[1], J. I. B. Camargo[9,2], R. Vieira-Martins[9,2], T. Swift[10], D. Dunham[10], T. George[10], J. Bardecker[10], C. Anderson[11,10], R. Nolthenius[16,10], K. Bender[16], G. Viscome[12,10], D. Oesper[10], R. Dunford[10], K. Getrost[10], C. Kitting[13], K. Green[14,15], R. Bria[10], A. Olsen[10], A. Scheck[10], B. Billard[17,10], M. E. Wasiuta[17], R. Tatum[10], P. Maley[10], D. di Cicco[18], D. Gamble[10], P. Ceravolo[19], D. Ceravolo[19], W. Hanna[10], N. Smith[10,20], N. Carlson[10], S. Messner[10,21], J. Bean[10], J. Moore[10] and R. Venable[10]

[1] *Universidade Federal do Rio de Janeiro - Observatório do Valongo, Ladeira do Pedro Antonio 43, Rio de Janeiro, RJ 20.080-090, Brazil*
[2] *Laboratório Interinstitucional de e-Astronomia - LIneA and INCT do e-Universo, Rua Gal. José Cristino 77, Rio de Janeiro, RJ 20921-400, Brazil*
[3] *Universidade Federal do Rio de Janeiro - CEDERJ, R. Antonio Ferreira 110 – Rio da Areia – Bacáxa, Saquarema, RJ 28990-000, Brazil*
[4] *LESIA, Observatoire de Paris - Section Meudon, 5 Place Jules Janssen - 92195 Meudon Cedex, France*
[5] *Universidade Federal de Uberlândia (UFU), Instituto de Física, Av. João Naves de Ávila 2121, Bairro Santa Mônica, Uberlândia, MG 38408-100, Brazil*
[6] *LESIA, Observatoire de Paris, Université PSL, Sorbonne Université, Université de Paris, CNRS, 92190 Meudon, France*
[7] *Federal University of Technology - Paraná (PPGFA/UTFPR-Curitiba), Av. Sete de Setembro, 3165, Curitiba, PR 80230-901, Brazil*
[8] *Universidade Estadual de So Paulo (UNESP), Grupo de Dinâmica Orbital e Planetologia, Guaratinguetá, SP 12516-410, Brazil*
[9] *Observatório Nacional - MCTI, R. General José Cristino 77, Rio de Janeiro, RJ 20.921-400, Brazil*
[10] *International Occultation Timing Association (IOTA), PO Box 20313, Fountain Hills, AZ 85269, USA*
[11] *Centennial Observatory, Herrett Center for Arts & Science, College of Southern Idaho, Twin Falls, Idaho, USA*
[12] *Rand Observatory II - MPC W71, Lake Placid, NY, USA*
[13] *Biological Sciences, California State University, Hayward, CA 94542, USA*
[14] *University of New Haven, 300 Boston Post Road, West Haven, CT 06516, USA*
[15] *Westport Astronomical Society, 182 Bayberry Lane, Westport, CT 06880, USA*
[16] *Cabrillo College and Earth Futures Institute, UC Santa Cruz, 6500 Soquel Drive, Aptos, CA 95003, USA*
[17] *Mark Slade Remote Observatory, Wilderness, Virginia, USA*
[18] *Sky & Telescope (Senior Contributing Editor), AAS Sky Publishing, LLC, USA*
[19] *Anarchist Mt. Observatory, British Columbia, Canada*
[20] *Barnard Astronomical Society, Chattanooga, Hamilton County, Harrison Bay State Park, TN, USA*
[21] *Harvest Moon Observatory, H25, Northfield, MN, USA*

**ABSTRACT**
We report the results of the stellar occultation by (UII) Umbriel on September 21st, 2020. The shadow crossed the USA and Canada, and 19 positive chords were obtained. A limb parameter accounted for putative topographic features in the limb fittings. Ellipse fittings were not robust – only upper limits were derived for the true size/shape of a putative Umbriel ellipsoid. The adopted spherical solution gives radius = 582.4 ± 0.8 km, smaller/close to 584.7 ± 2.8 km from Voyager II. The apparent ellipse fit results in a true semi-major axis of 584.9 ± 3.8 km, semi-minor axes of 582.3 ± 0.6 km and true oblateness of 0.004 ± 0.008 for a putative ellipsoid. The geometric albedo was $p_V = 0.26 \pm 0.01$. The density was $\rho = 1.54 \pm 0.04$ g cm$^{-3}$. The surface gravity was 0.251 ± 0.006 m s$^{-2}$ and the escape velocity 0.541 ± 0.006 km s$^{-1}$. Upper limits of 13 and 72 nbar (at $1\sigma$ and $3\sigma$ levels, respectively) were obtained for the surface pressure of a putative isothermal CO$_2$ atmosphere at $T$ = 70 K. A milliarcsecond precision position was derived: $\alpha = 02^h\ 30^m\ 28.^s84556 \pm 0.1$ mas, $\delta = 14^o\ 19'\ 36''.5836 \pm 0.2$ mas. A large limb parameter of 4.2 km was obtained, in striking agreement with opposite southern hemisphere measurements by Voyager II in 1986. Occultation and Voyager results indicate that the same strong topography variation in the surface of Umbriel is present on both hemispheres.

# 1 INTRODUCTION

The Uranian moon system has been extensively explored from dynamical and astrometrical points of view. The origin of Uranus's spin in connection with the orbital architecture of the moons is being investigated (Saillenfest et al. 2022; Rufu & Canup 2022), and even a concept space mission has been proposed to measure the spin orientation/inclination with precision for mitigating systematic bias in dynamical studies (Iorio, Girija & Durante 2023). Orbital theory for the moons has evolved significantly since the 80s (Laskar & Jacobson 1987; Lainey 2016), and good quality ephemerides have been produced (e.g. Jacobson 2014; Emelyanov, Varfolomeev & Lainey 2022). Ephemeris work has been fed by space astrometry with Voyager II (Jacobson 1992a) and the Hubble Space Telescope (Showalter & Lissauer 2006), by long-term ground-based ($\alpha, \delta$) astrometry programs (Veiga, Martins & Andrei 2003; Qiao et al. 2013; Xie et al. 2019; Camargo et al. 2022; Zhang et al. 2022), one mutual phenomena campaign (Christou et al. 2009; Mallama et al. 2009; Assafin et al. 2009; Arlot et al. 2013) and two (UIII) Titania stellar occultations (Widemann et al. 2009). Also, alternative astrometry methods have been proposed/applied without the use of astrometric catalogues, as Uranus is currently crossing a sky region without sufficient reference stars (Santos-Filho et al. 2019).

★ E-mail: massaf@ov.ufrj.br

On the other hand, astrophysical studies for the Uranus system are relatively sparse in comparison to other Solar System giant planets. Uranus presents a rich system with five large moons with interesting surface features and the potential to harbor oceans (Cochrane et al. 2021), besides having thirteen known moonlets and a complex ring system (Showalter 2020). It is thought from integrated disk spectra that the main moons are formed of $H_2O$ ice, carbon-based ices including $CO_2$, and nitrogen-based species such as $NH_3$, which suggests recent geological activity, but the true composition and distribution of the moons' constituents (including the presence of organics) is essentially unknown (Cartwright & Beddingfield 2022). Having heavier "ices" (water, methane, hydrogen sulfide, ammonia) than the hydrogen which dominates Jupiter and Saturn composition (Cohen et al. 2022), Uranus- and Neptune-type ice giants may be typical representatives of a common class of exoplanets (Batalha et al. 2011; Wakeford & Dalba 2020). The Uranus system is thus a very interesting laboratory for the astrophysical study of this peculiar and important class of ice giants.

Uranus (together with Neptune) is comparatively the least explored of the planets in our Solar System. Only once has it been (briefly) visited by a space probe. Voyager II observed the illuminated southern hemisphere of Uranus and its moons in 1986 (Thomas 1988). A launch window opportunity will occur in 2030–2034 due to a favorable gravity assist configuration by Jupiter, allowing a probe to reach the system before its 2050 equinox, after which the northern hemisphere will again become inaccessible. This motivated the proposal to send the first orbiters to investigate the Uranus system (Cohen et al. 2022; Cartwright & Beddingfield 2022). These missions this time would benefit from magnetic field and infrared detectors among other instruments, allowing for a much better understanding of the composition, including the detection of organics and the discovery of putative subsurface oceans in the moons. The comparison between northern and southern hemisphere results is also essential to this investigation. An adequate preparation for such a challenging space mission involves the production and maintenance of precise/accurate ephemerides for equipment optimization (more science cargo, less fuel) and efficient navigation, and as much precise knowledge as possible of basic physical properties of targets such as size, shape and brightness, for a successful planning and optimization of in situ observations.

Stellar occultations by natural satellites observed from the ground (Morgado et al. 2022) may provide sizes with kilometer accuracy/precision, rivaling space probe measurements. Natural moons' sizes, shapes, albedos, and densities measured with this technique furnish fundamental astrophysical information, which can also be used to optimize space missions. Putative atmospheres can be also probed. Ephemeris work also benefits from milliarcsecond (mas) precision positions obtained by the technique. However, so far only two observed events have been published for the Uranus system: the 2001 and 2003 stellar occultations by Titania (Widemann et al. 2009).

Here, we report the results of the occultation of a $m_V$ = 13.4 star by Umbriel observed on 2020 September 21 in the USA and Canada. For years, Uranus has been crossing a region of sky with very few stars, so this event provided a rare opportunity to probe Umbriel's northern hemisphere, now accessible. From the campaign involving 29 participants, 19 positive chords could be derived. In Section 2, we describe the observations. Section 3 describes the photometry and Section 4 details the light curve fittings to derive the ingress/egress occultation instants of the chords. In Section 5, we describe the limb fittings of the chords. The results are presented in Section 6, where we determine Umbriel's size, shape, density, and albedo, among other physical characteristics. Atmosphere limits are also estimated, besides a milliarcsecond-precision position as a by-product of the fittings. In Section 7, we discuss the results in the context of Voyager II measurements and present our conclusions.

## 2 OBSERVATIONS

The occultation was first predicted and posted by the International Occultation Timing Association[1] (IOTA). Umbriel occulted the magnitude 13.4 star Gaia DR3 75195604519240064 on 2020 September 21 at the reference instant 8 h 24 m 36s (UTC). The occultation shadow crossed the USA and Canada with a geocentric shadow velocity of -17.12 km s$^{-1}$. Table 1 furnishes detailed information about the star and Umbriel at the occultation reference instant. Of 29 participants, 23 were inside and six outside the shadow. Among the inside sites, two were overcast, one had technical problems and one observation set was unfortunately lost. Among the outside sites, one was overcast. In all, 19 positive and five negative chords were derived. No filters were used to improve the signal-to-noise ratio. Most observations were made by IOTA members. Most observers were experienced amateurs, with the participation of professional astronomers and professors too, all guided by occultation campaign tools such as Occult[2], OccultWatcher (OW)[3] (Pavlov 2018a,b) and its online version OccultWatcher Cloud[4] (OWC). The 10 – 60 cm aperture telescopes employed were equipped with CCD or video camera detectors. Time was registered with GPS devices or by internet NTP synchronization. The video recording in `ser`[5], `avi` and `mov` formats were converted to `FITS` images (Wells et al. 1981) by using `Tangra` version 3.7.3 software[6], and proper care was taken to correctly retrieve the image instants – see details in Benedetti-Rossi et al. (2016) and Barry et al. (2015). Table 2 summarizes the location, instruments and observation status of the campaign participants. Fig. 1 displays the shadow path of the occultation on Earth.

---

[1] IOTA prediction by T. J. Swift and A. R. Gomes Júnior: https://www.occultationpages.com/events/20200921_Umbriel.html
[2] Occult: www.lunar-occultations.com/iota/occult4.htm
[3] OccultWatcher (OW): https://www.occultwatcher.net/
[4] OccultWatcher Cloud: https://cloud.occultwatcher.net/
[5] Videos in ser format: http://www.grischa-hahn.homepage.t-online.de/astro/ser/
[6] `Tangra` software: http://www.hristopavlov.net/Tangra3/

**Table 1.** Star, Umbriel and Uranus information at the occultation.

| | Occulted star |
|---|---|
| Epoch | 2020-09-21 08:24:36.000 UTC |
| Source ID | Gaia DR3 75195604519240064 |
| Star position [1] | $\alpha = 02^h\ 30^m\ 28.^s84657 \pm 0.0873$ mas |
| | $\delta = 14^o\ 19'\ 36''.3762 \pm 0.0860$ mas |
| Magnitudes [2] | R = 13.474 ± 0.002; G = 13.779 ± 0.002 |
| | B = 14.183 ± 0.002; J = 11.904 ± 0.021 |
| | H = 11.412 ± 0.023; K = 11.323 ± 0.022 |
| | Absolute G magnitude = 3.214443 |
| RUWE [3] | 1.11 |
| Apparent diameter [4] | 0.0246 mas = 0.339 km |
| Star class [5] | Main sequence dwarf or sub-giant G-type star |
| | Umbriel |
| Ephemeris [6] | DE435/URA111 |
| Geocentric distance | 19.0194591323518 au |
| Apparent velocity | 17.2 km s$^{-1}$ (relative velocity Umbriel - star) |
| Apparent magnitude [7] | V = 14.978 |
| Mass [8] | $(1.275 \pm 0.028) \times 10^{21}$ kg |
| Rotation period [9] | 4.144 days |
| Pole [10] | RA = $17^h\ 08^m\ 55.^s6624$ |
| | Dec = $-15^o\ 02'\ 00''.809$ |
| Position angle [11] | P = 145.082 deg |
| Aspect angle [12] | $\zeta = 38^o\ 16'\ 24''.82$ (north pole) |
| Sub-observer point [13] | Longitude = $204^o\ 20'\ 04''.518$ |
| at Umbriel | Latitude = $+51^o\ 43'\ 35''.183$ |
| Sub-Uranus point [14] | Longitude = $1^o\ 04'\ 17''.394$ |
| at Umbriel | Latitude = $-0^o\ 11'\ 08''.356$ |

*Notes.* (1) Star position taken from the Gaia Data Release 3 (GDR3) star catalog (Gaia Collaboration 2023), propagated to the event epoch following Butkevich & Lindegren (2014) with proper motion corrections by Cantat-Gaudin & Brandt (2021), using SORA (Gomes-Júnior et al. 2022). (2) R, G, B magnitudes from Gaia DR3 BP and RP spectrophotometry (Carrasco et al. 2023); J, H, K from 2MASS catalogue (Cutri et al. 2003); StarHorse absolute G magnitude from Anders et al. (2019); all magnitude values retrieved with Vizier service (http://vizier.cds.unistra.fr/viz-bin/VizieR-4). (3) Renormalized unit weight error (Lindegren et al. 2021). (4) Angular diameter from Gaia DR2 (Andrae et al. 2018). (5) G spectral type from GDR3; star class is discussed in Section 4. (6) Extracted with SORA by querying the JPL Horizons web service (https://ssd.jpl.nasa.gov/horizons/app.html#/). (7) Visual magnitude from the JPL Horizons web service. (8) Taken from Jacobson (2014). (9) Umbriel is synchronous with Uranus (Smith et al. 1986); period from Archinal et al. (2018). (10) Rotation axis' pole from Archinal et al. (2018). (11) Apparent position angle *P* of Umbriel's putative ellipsoid projected in the sky plane, taken with respect to the semi-minor axis of the apparent ellipse in the sky plane (north = $0^o$ and increasing counterclockwise to the east); value taken considering the satellite-Uranus geometry (see Section 6.1). (12) The angle between the line of sight and the rotation axis's north pole of Umbriel. (13) Sub-point coordinates of the observer (geocenter) at the Umbriel-centered coordinate system (Archinal et al. 2018). (14) Sub-point coordinates of Uranus at the Umbriel-centered coordinate system. Aspect angle, observer and Uranus sub-point coordinates computed using DE435/URA111 and Archinal et al. (2018). Values at occultation epoch.

## 3 LIGHT CURVE PHOTOMETRY

Occultation light curves for each observation were derived from differential aperture photometry. In some cases, digital coronagraphy was necessary prior to the photometry to remove light contamination from Uranus. Digital coronagraphy and photometry were performed with the corresponding tools of the PRAIA package (Assafin 2023). The flux of the occulted star mixed with Umbriel (target flux) was calibrated by the flux of nearby field stars and satellites Titania and Oberon to account for sky transparency variations - see a typical FOV sample in Fig. 2. The target/calibration flux ratio was normalized to the full flux ratio outside the occultation. PRAIA automatically sampled calibrators and the set resulting in the minimum flux ratio dispersion outside the occultation was used. In cases where no useful reference object was available, the light curve was obtained from uncalibrated Umbriel flux, or using Uranus as calibrator - whichever gave the least flux dispersion. The di Cicco video recording was lost, so no curve could be derived from it. The 19 positive light curves are plotted in Fig. 3. The normalized flux ratio dispersion (standard deviation) outside the occultation is given in Table 3 as a light curve photometric precision indicator – more details about Table 3 are given in Section 4.

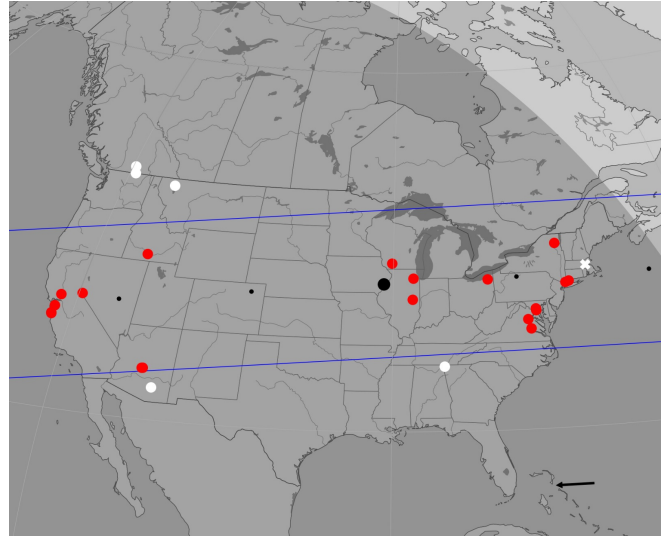

**Figure 1.** Occultation map with the fitted shadow path crossing the USA and Canada. Sites with measured positive chords are red, negative ones white. The white X inside the shadow did not produce a chord (see text and Table 2). Black dots are spaced by 1 minute and the biggest one represents the closest geocenter approach at the reference occultation instant 08h 24m 36s (event date). The arrow indicates the direction of the shadow's motion.

## 4 INGRESS AND EGRESS OCCULTATION INSTANTS

Assuming an opaque body without atmosphere, the ingress and egress times of all chords were calculated by modeling the light curves with a sharp-edge (box geometry) occultation model convolved with Fresnel diffraction, the apparent stellar diameter at the object's distance, and the finite exposure time of each observation set with the SORA package (Gomes-Júnior et al. 2022).

Gaia DR2 classified the occultation star as G-type in the MK system. For V ≡ G, an absolute V magnitude = 3.214443 (Anders et al. 2019) and index B-V = 0.404 ± 0.003 (see Table 1) place the star as a main sequence dwarf or sub-giant star in the H-R diagram. Following the empirical models based in B, V (≡ G) and K magnitudes by Kervella (2004) for dwarfs and sub-giants, and by van Belle (1999) for main sequence stars, we compute limb-darkening-corrected apparent diameters of 0.0254 and 0.0247 mas for the star, respectively. Limb-darkening was computed by fitting the spectral energy distribution of

**Table 2.** Observational information.

| Observer | Latitude (° ′ ″), Longitude (° ′ ″), Altitude (m) | D(cm), f/, Detector, Format | Observation (UTC) Start (08:mm:ss) End (08:mm:ss) | Exp. time (s), Cycle time (s), Time device | Status |
|---|---|---|---|---|---|
| Dave Gamble, CAN | +49 35 34.000, -119 41 55.900, 522 | 45.70, 04.5, 1, FITS | 19:00 - 29:00 | 0.11300, 0.11300, GPS | N |
| Peter & Debra Ceravolo, CAN | +49 00 32.000, -119 21 47.000, 1097 | 28.00, 05.0, 1, FITS | 24:00 - 28:00 | 0.35000, 0.35141, GPS | N |
| Willian Hanna, USA | +48 23 38.730, -114 12 43.960, 979 | 27.94, 10.0, 1, FITS | 21:37 - 30:03 | 1.00000, 1.00235, GPS | N |
| Chris Anderson, USA | +42 35 01.800, -114 28 13.200, 1120 | 60.00, 12.8, 4, AVI | 22:48 - 30:15 | 2.06832, 2.06930, GPS | P |
| Steve Messner, USA | +44 29 57.500, -093 07 45.080, 289 | 45.00, 03.0, 3 | | | O |
| George Viscome, USA | +44 15 40.140, -074 00 25.460, 596 | 36.83, 06.0, 2, FITS | 19:31 - 25:38 | 0.17500, 0.17630, GPS | P |
| David Oesper, USA | +42 57 36.900, -090 08 31.100, 390 | 30.50, 03.3, 3, AVI | 09:00 - 36:10 | 0.12800, 0.13310, GPS | P |
| James Bean, USA | | | | | O |
| Dennis di Cicco, USA | +42 21 01.000, -071 23 21.000, 43 | 40.60, 20.0, 5 | 20:01 - 28:00 | 0.25000, 0.25000, NTP | L |
| Jerry Bardecker, USA | +38 53 23.500, -119 40 20.300, 1524 | 30.48, 10.0, 3, AVI | 23:15 - 29:30 | 0.03330, 0.03330, GPS | P |
| Theodore J. Swift, USA | +38 33 08.260, -121 47 08.140, 18 | 20.00, 10.0, 3, AVI | 24:30 - 28:30 | 0.26400, 0.26700, GPS | P |
| Robert Dunford, USA | +41 45 32.400, -088 07 00.010, 230 | 35.56, 01.9, 1, FITS | 18:52 - 28:46 | 3.00000, 3.00070, GPS | P |
| Kai Getrost, USA | +41 35 06.100, -081 04 45.720, 348 | 25.40, 10.0, 1, FITS | 20:01 - 28:06 | 0.50000, 0.50030, GPS | P |
| Chris Kitting, USA | +37 38 48.840, -122 02 09.096, 189 | 25.40, 04.7, 3, AVI | 23:00 - 30:30 | 0.26640, 0.26670, GPS | P |
| Kevin Green, USA | +41 10 15.900, -073 19 39.300, 87 | 35.60, 07.7, 1, SER | 18:39 - 27:18 | 1.00000, 1.00220, GPS | P |
| Rick Bria, USA | +41 04 01.000, -073 41 30.000, 118 | 35.50, 07.2, 3, AVI | 20:00 - 26:05 | 0.13320, 0.13320, GPS | P |
| Kirk Bender, USA | +37 03 27.470, -122 07 23.200, 555 | 20.32, 10.0, 3, AVI | 23:30 - 31:00 | 0.26640, 0.26700, GPS | P |
| Richard Nolthenius, USA | +37 01 04.120, -122 04 45.310, 341 | 20.32, 06.3, 3, AVI | 23:10 - 30:00 | 0.52800, 0.53390, GPS | P |
| Aart Olsen, USA | +40 05 12.400, -088 11 46.300, 224 | 50.00, 04.0, 6, AVI | 19:27 - 29:27 | 0.40000, 0.40000, GPS | P |
| Andrew Scheck, USA | +39 08 59.070, -076 53 13.330, 120 | 20.00, 06.3, 7, MOV | 20:36 - 24:51 | 0.99000, 1.00050, GPS | P |
| David Dunham, USA Barton Billard & | +38 59 12.745, -076 52 08.880, 46 | 41.00, 04.4, 1, FITS | 20:01 - 25:04 | 1.00000, 1.00050, GPS | P |
| Myron E. Wasiuta, USA | +38 20 02.000, -077 42 38.000, 96 | 10.20, 07.0, 1, FITS | 20:51 - 26:51 | 4.00000, 4.00172, GPS | P |
| John Moore, USA | | | | | T |
| Randy Tatum, USA | +37 35 42.576, -077 33 02.484, 77 | 30.00, 10.0, 8, AVI | 21:22 - 26:16 | 0.10000, 0.10000, NTP | P |
| Paul Maley, USA | +33 48 42.858, -111 57 07.974, 654 | 28.00, 05.0, 3, AVI | 24:57 - 27:55 | 0.53280, 0.53390, GPS | P |
| Tony George, USA | +33 49 00.100, -111 52 07.300, 843 | 30.00, 03.3, 3, AVI | 24:30 - 28:30 | 0.13320, 0.13390, GPS | P |
| Ned Smith, USA | +34 52 30.000, -085 28 15.600, 210 | 63.50, 03.2, 1, FITS | 20:14 - 28:13 | 1.00000, 1.00011, GPS | N |
| Normam Carlson, USA | +32 25 53.140, -110 44 43.940, 2391 | 23.50, 00.5, 9 AVI | 23:28 - 30:00 | 0.13340, 0.13340, NTP | N |
| Roger Venable, USA | +32 22 25.080, -083 12 06.720, 103 | 35.60, 07.7, 3 | | | O |

*Notes.* D(cm) = diameter. f/ = focal ratio. Status: P = positive, N = negative, L = video recording lost, T = technical problems, O = overcast. Detectors: 1 = QHY 174, 2 = QHY 174M, 3 = Watec 910HX, 4 = Watec 120N+, 5 = Celestron Skyris 618M, 6 = Watec 910BD, 7 = Mallincam, 8 = ZWO ASI224MC, 9 = RunCam Night Eagle Astro. Chords ordered from north to south in the sky plane.

the star with Castelli & Kurucz (2003) atmosphere models, using the reddening definition by Fitzpatrick (1999). These values are close to each other and to the apparent stellar diameter of 0.0246 mas = 0.339 km computed from Gaia DR2 data (Andrae et al. 2018), used as the apparent stellar diameter in our light curve fittings (Table 1). The Fresnel scale at D = 19.02 au (astronomical units) is $L_f = \sqrt{\lambda D/2}$ = 1.0 km for a typical wavelength of $\lambda$ = 0.65 µm. The apparent velocity of the event was -17.12 km s$^{-1}$. Since only 5 observations were made with exposure times less than about 0.1 s = 1.72 km spatial resolution, the behaviour of the light curves were generally dominated by the exposure time, not by Fresnel diffraction or star diameter.

For a few chords far from the body's center, we refined the input velocity values by fitting the local normal velocity, taking into account the estimated angular diameter of the star, wavelength and bandwidth of each observation. We then updated their input event velocities and repeated the light curve fittings to determine the final instants. The instants did not change significantly, so we discarded these test values and used the nominal velocity $v$ = -17.12 km s$^{-1}$ for all chords.

Prior to the di Cicco video loss, visual inspection yielded $T_{IN}$ = 8h 22m 17.5 s and $T_{EG}$ = 8h 23m 02.5 s ± 0.6 s for the ingress/egress instants. Since the error bars are large in comparison to the error values of close-fitted chords and the instants cannot be effectively verified, we conservatively discarded this result.

Table 3 lists the duration/length, ingress and egress time instants (and errors) for all 19 positive chords fitted with SORA. Furnished in advance, the (O-C)$_{IN}$ (ingress) and (O-C)$_{EG}$ (egress) offsets listed in Table 3 are the radial distances between the chord extremities and the fitted circle/ellipse limbs, with positive/negative values for extremities outside/inside the limb (see Section 5). Photometric information for each chord is also given. A typical example of light curve fitting is displayed in Fig. 4.

## 5 LIMB FITTINGS

The event provided 19 positive chords uniformly spread across Umbriel, and five negatives with the nearest ones at 293.0 and 93.2 km from the limb (after fitting), respectively at the northern and southern sides of the body's figure in the sky plane. We used SORA to fit a circle and an ellipse to the chords in the sky plane. For a circular limb, the $M$ = 3 fitted parameters are the observed $f_c$ and $g_c$ geocentric ephemeris offsets projected in the sky plane respectively in $\alpha$ and $\delta$, and the radius $R$. For an elliptical limb, we fit $M$ = 5 parameters. Besides $f_c$ and $g_c$, we have the apparent semi-major axis

Table 3. Ingress/egress occultation instants and light curve photometry for the 19 positive chords.

| Observer | Ingress (08:mm:ss.s) | σ(s) σ(km) | Egress (08:mm:ss.s) | σ(s) σ(km) | Chord length (km) | (O-C)$_{IN}$ (km) Circle Ellipse | (O-C)$_{EG}$ (km) Circle Ellipse | σ (flux) | Coro Cal |
|---|---|---|---|---|---|---|---|---|---|
| Anderson | 25:51.712 | 0.943 16.220 | 26:46.677 | 0.357 6.140 | 945.398 ± 17.343 | +18.346 ± 16.439 19.533 | -5.560 ± 6.225 -5.722 | 0.13 | N N |
| Viscome | 22:45.657 | 0.028 0.482 | 23:46.549 | 0.023 0.396 | 1047.342 ± 0.623 | +3.662 ± 0.488 +4.602 | +4.105 ± 0.401 +3.940 | 0.10 | N C |
| Oesper | 23:57.557 | 0.022 0.378 | 25:01.655 | 0.024 0.413 | 1102.486 ± 0.560 | +1.833 ± 0.384 +2.502 | -3.412 ± 0.419 -3.489 | 0.11 | C C |
| Bardecker | 26:17.330 | 0.066 1.135 | 27:23.519 | 0.069 1.187 | 1138.451 ± 1.642 | -2.417 ± 1.151 -2.034 | -1.557 ± 1.203 -1.460 | 0.11 | C C |
| Swift | 26:26.857 | 0.187 3.216 | 27:33.196 | 0.065 1.118 | 1141.031 ± 3.405 | +0.217 ± 3.260 +0.597 | -2.140 ± 1.133 -2.040 | 0.14 | C C |
| Dunford | 23:47.455 | 0.373 6.416 | 24:54.651 | 0.385 6.622 | 1155.771 ± 9.220 | -2.138 ± 6.724 -1.846 | +12.690 ± 2.970 +12.852 | 0.10 | C C |
| Getrost | 23:14.272 | 0.047 0.808 | 24:21.048 | 0.047 0.808 | 1148.547 ± 1.143 | -7.269 ± 0.821 -7.086 | +8.724 ± 0.821 +8.976 | 0.09 | C C |
| Kitting | 26:29.916 | 0.047 0.808 | 27:36.347 | 0.051 0.877 | 1142.613 ± 1.193 | +0.923 ± 0.819 +1.073 | -6.404 ± 0.889 -6.113 | 0.15 | C C |
| Green | 22:37.692 | 0.085 1.462 | 23:43.770 | 0.053 0.912 | 1136.542 ± 1.723 | -5.969 ± 1.483 -5.904 | -0.355 ± 0.925 +0.010 | 0.11 | C C |
| Bria | 22:39.130 | 0.031 0.533 | 23:45.047 | 0.038 0.654 | 1133.772 ± 0.844 | -1.692 ± 0.541 -1.650 | -5.146 ± 0.663 -4.751 | 0.18 | C C |
| Bender | 26:31.960 | 0.133 2.288 | 27:38.105 | 0.073 1.256 | 1137.694 ± 2.610 | +1.211 ± 2.319 +1.226 | -1.633 ± 1.273 -1.198 | 0.19 | N C |
| Nolthenius | 26:31.794 | 0.206 3.543 | 27:37.742 | 0.180 3.096 | 1134.306 ± 4.705 | +3.326 ± 3.592 +3.332 | -5.914 ± 3.140 -5.464 | 0.35 | C C |
| Olsen | 23:49.104 | 0.092 1.582 | 24:53.865 | 0.118 2.030 | 1113.889 ± 2.574 | -1.102 ± 1.607 -1.206 | -0.412 ± 2.062 +0.170 | 0.10 | C C |
| Scheck | 22:56.083 | 0.283 4.868 | 23:56.284 | 0.214 3.681 | 1035.457 ± 6.103 | +3.751 ± 3.943 +3.484 | +2.744 ± 3.739 +3.660 | 0.07 | N C |
| Dunham | 22:55.982 | 0.210 3.612 | 23:55.951 | 0.202 3.474 | 1031.467 ± 5.012 | +10.195 ± 3.668 +9.919 | +7.622 ± 3.529 +8.576 | 0.14 | C C |
| Billard-Wasiuta | 23:03.893 | 0.711 12.229 | 23:57.896 | 0.414 7.121 | 928.852 ± 14.151 | -15.204 ± 12.424 -15.484 | +13.803 ± 7.235 +14.912 | 0.08 | C N |
| Tatum | 23:05.211 | 0.030 0.516 | 23:51.710 | 0.033 0.568 | 799.783 ± 0.767 | +2.691 ± 0.524 +2.481 | -5.068 ± 0.577 -3.785 | 0.25 | N C |
| Maley | 26:14.879 | 0.073 1.256 | 26:35.260 | 0.099 1.703 | 350.553 ± 2.116 | +0.606 ± 1.276 +0.974 | -1.399 ± 1.731 -0.163 | 0.09 | C C |
| George | 26:14.606 | 0.018 0.310 | 26:34.791 | 0.020 0.344 | 347.182 ± 0.463 | +0.461 ± 0.315 +0.836 | -1.078 ± 0.350 +0.156 | 0.09 | C C |

*Notes.* Coronagraphy (Coro): C and N mean coronagraphed or not. Photometric calibration (Cal): C and N mean calibrated or not. Normalized light curve flux ratio dispersion σ(flux) is the standard deviation computed outside of the occultation. The (O-C)$_{IN}$ (ingress) and (O-C)$_{EG}$ (egress) offsets are the radial distance differences between the measured chord extremities and the fitted circular/ellipse limbs, with positive/negative values for extremities outside/inside the limbs (see Section 5). The (O-C) errors are the measured chord errors in the extremities projected in the radial directions; they are the same for circular/elliptical limbs. Chords are ordered from north to south in the sky plane.

$a'$, the apparent oblateness $\epsilon' = (a' - b')/a'$ where $b'$ is the apparent semi-minor axis, and the position angle $P$ of $b'$ in the sky plane (with north = $0^o$, increasing counterclockwise to the east). For $N$ = 38 chord extremities projected in the sky plane, we have $N - M = 35$ and $N - M = 33$ degrees of freedom for circular and elliptical fittings respectively. We also input a limb parameter in the fittings – SORA's $\sigma_{model}$ parameter of equation 11 in Gomes-Júnior et al. (2022). It regulates the uncertainty of the circle/ellipse limbs associated with the putative existence of topographic features on the surface of the body (see also Rommel et al. 2023). This is justified as Umbriel – along with Miranda – has the most irregular limb among the Uranian satellites (Thomas 1988; Schenk & Moore 2020). The best $\chi^2$ for the circular fitting – expected to be around $N/(N-M)=38/(38-3) = 1.086$ – was achieved with $\sigma_{model}$ = 4.2 km. The position angle could not be found without large uncertainties in the elliptical fitting, so we fixed it using the expected value (Table 1) from the satellite-Uranus geometry (see Section 6.1). The degrees of freedom for the elliptical fitting was thus $N - M = 34$, and the best $\chi^2$ – expected to be around $N/(N-M)=38/(38-4) = 1.118$ – was achieved with $\sigma_{model}$ = 4.3 km. Table 4 displays the results for the circular and elliptical limb fittings. Fig. 5 displays the chords fitted with a circular limb, visually indistinguishable from the fitted elliptical one.

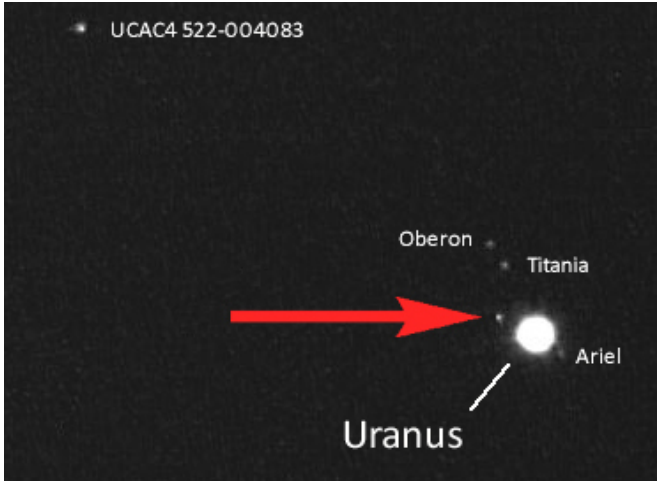

**Figure 2.** Cut FOV from site Bardecker with ∼ 4.3' x 3.2' (original size 5.5' x 3.75'). North up, east left. The top/left object is a magnitude G = 16.8 nearby field star (UCAC4 522-004083 = GDR3 75192683941477888) used as photometric calibrator. Except for Miranda, all main satellites are visible. Titania and Oberon were also used as calibrators. The target (occultation star mixed with Umbriel) is indicated by the red arrow.

**Table 4.** Umbriel circular and elliptical limb fittings of the 19 positive chords.

| Circular limb | |
|---|---|
| Radius | $R = 582.4 \pm 0.8$ km (1-sigma) |
| Ephemeris offsets | $f_c = -23.7 \pm 0.8$ km |
|  | $g_c = -169.4 \pm 1.9$ km |
| Limb parameter | $\sigma_{model} = 4.2$ km |
| $\chi^2$ per degree of freedom | 1.077 |
| Elliptical limb | |
| Position angle [1] | P = 145.082 deg (fixed) |
| Apparent semi-major axis | $a' = 582.6 \pm 1.1$ km (1-sigma) |
| Apparent oblateness | $\epsilon' = 0.003 \pm 0.003$ (1-sigma) |
| Apparent equivalent radius [2] | $R'_{eq} = 581.7 \pm 2.0$ km |
| Ephemeris offsets [3] | $f_c = -23.7 \pm 0.9$ km |
|  | $g_c = -169.5 \pm 2.0$ km |
| Limb parameter | $\sigma_{model} = 4.3$ km |
| $\chi^2$ per degree of freedom | 1.104 |

*Notes.* (1) The position angle was fixed by the expected value (Table 1) from the satellite-Uranus geometry (see Section 6.1). (2) The apparent equivalent radius is the radius of a circle with the same area of the fitted apparent ellipse, i.e. $R'_{eq} = a'\sqrt{1-\epsilon'}$. (3) Ephemeris offsets in the sense "Occultation minus DE435/URA111".

## 6 RESULTS

### 6.1 Sizes and shapes

A body of radius about 580 km such as Umbriel is within the critical minimum size ranges required for icy bodies (200–900 km) or rocky ones (500–1200 km) to be in hydrostatic equilibrium (Chandrasekhar 1987; Tancredi & Favre 2008). Since it is synchronous and tidally locked with Uranus, having an extremely slow ≈ 4.1 day rotation period (Table 1), Umbriel is believed to have an ellipsoid shape, with the major axis always pointing at Uranus and the rotation axis perpendicular to its orbit around the planet (Karkoschka 2001). Since Umbriel has a very small Δmag = 0.02 ± 0.04 brightness variation

(Karkoschka 2001), the other axes should have about the same size and not differ too much from the length of the major axis, such as with Miranda and Ariel (Thomas 1988). As the aspect angle of the north rotation pole with respect to the occultation observers was about $38^o$ (Table 1), Umbriel's shape projected in the sky plane should render an elliptical limb.

Following the rationale by Dermott & Thomas (1988), we can determine the axial lengths of Umbriel from the fitted apparent elliptical limb and the geometry of the observation. Using the right-handed Umbriel-centered coordinate frame with *x* and *z* axes pointing respectively at Uranus and the north rotation pole (see Fig. 6), we can relate the fitted apparent semi-major axis $a'$ and oblateness $\epsilon'$ (and errors) with the true semi-major axis $a$ and semi-minor axes $b = c$ (and errors), using formulae 8 - 12 in Dermott & Thomas (1988). The exact expressions in equation 1 use auxiliary quantities $r$, $s$ and $t$. The two known angles $\psi$ and $\varphi$ give the direction of the occultation observers defined in the Umbriel-centered coordinate frame (Fig. 6). The angle $\varphi$ is the difference between the sub-observer and sub-Uranus longitude points and $\psi$ is the sub-observer latitude, given in Table 1 at occultation epoch.

$$\frac{1}{a'^2} = \frac{r + s^{1/2}}{2t}, \quad \frac{1}{c'^2} = \frac{r - s^{1/2}}{2t}$$
$$r = (c^2 - a^2)\cos^2\psi\,\cos^2\varphi + a^2 + c^2$$
$$t = (c^4 - a^2c^2)\cos^2\psi\,\cos^2\varphi + a^2c^2$$
$$s = a^4(1 - \cos^2\psi\,\cos^2\varphi)^2 +$$
$$c^4\cos^2\psi(\cos^2\psi + 2\sin^2\psi\,\sin^2\varphi - 2\cos^2\varphi) +$$
$$2a^2c^2(\cos^4\psi\,\sin^2\varphi\,\cos^2\varphi - \sin^2\psi) +$$
$$2a^2c^2\cos^2\psi(\sin^2\psi\,\cos^2\varphi - \sin^2\varphi) +$$
$$c^4(1 - \cos^2\psi\,\sin^2\varphi)^2 \quad (1)$$

We probed true semi-major and minor axes using a Uniform Random Deviate random number generator (Press et al. 1982). After apparent values $a'$ and $c' = a'\sqrt{1 - \epsilon'}$ computed with equation 1 match *N* times the fitted values within their 1-sigma errors (Table 4), we use the corresponding input true values to compute weighted means and 1-sigma weighted standard-deviations for $a$ and $c = a\sqrt{1-\epsilon}$. The (O-C)s between the generated and fitted apparent semi-major and minor axes are both used as weights for each true axis. For $N$ = 1 million, we get $a = 584.9 \pm 3.8$ and $c = b = 582.3 \pm 0.6$ km as the true ellipsoid semi-axes of Umbriel, with true oblateness $\epsilon = 0.004 \pm 0.008$ and equivalent radius $R_{eq} = 583.6 \pm 2.2$ km as the radius of a circle with the same area as an $(a,c)$ ellipse.

Since the major axis of Umbriel's ellipsoid points to Uranus, the apparent ellipse's major axis is aligned with the planet direction in the sky plane, resulting in an apparent position angle of $145^o.082$ during occultation (Table 1). We fixed this value to improve the degrees of freedom in the ellipse limb fitting.

However, treating the apparent position angle as a free parameter, we could not fit values without large uncertainties. We tested the robustness of the ellipse fitting by fixing fake apparent position angles in the range ± $90^o$ around the correct value, and using a fixed limb parameter of $\sigma_{model} = 4.3$ km. The $\chi^2$ variation of the fits did not exceed ± 0.004, indicating that the data was not sensitive to the elliptical limb model. This conclusion is also supported by the obtained true oblateness, which is not statistically significant even at the 1-sigma level.

Therefore, we take the obtained $a$, $c$ and $\epsilon$ values as upper limits for the true semi-major/minor axes and oblateness of a putative ellipsoid shape for Umbriel, and assume the spherical solution as the main result with a fitted radius of 582.4 ± 0.8 km.

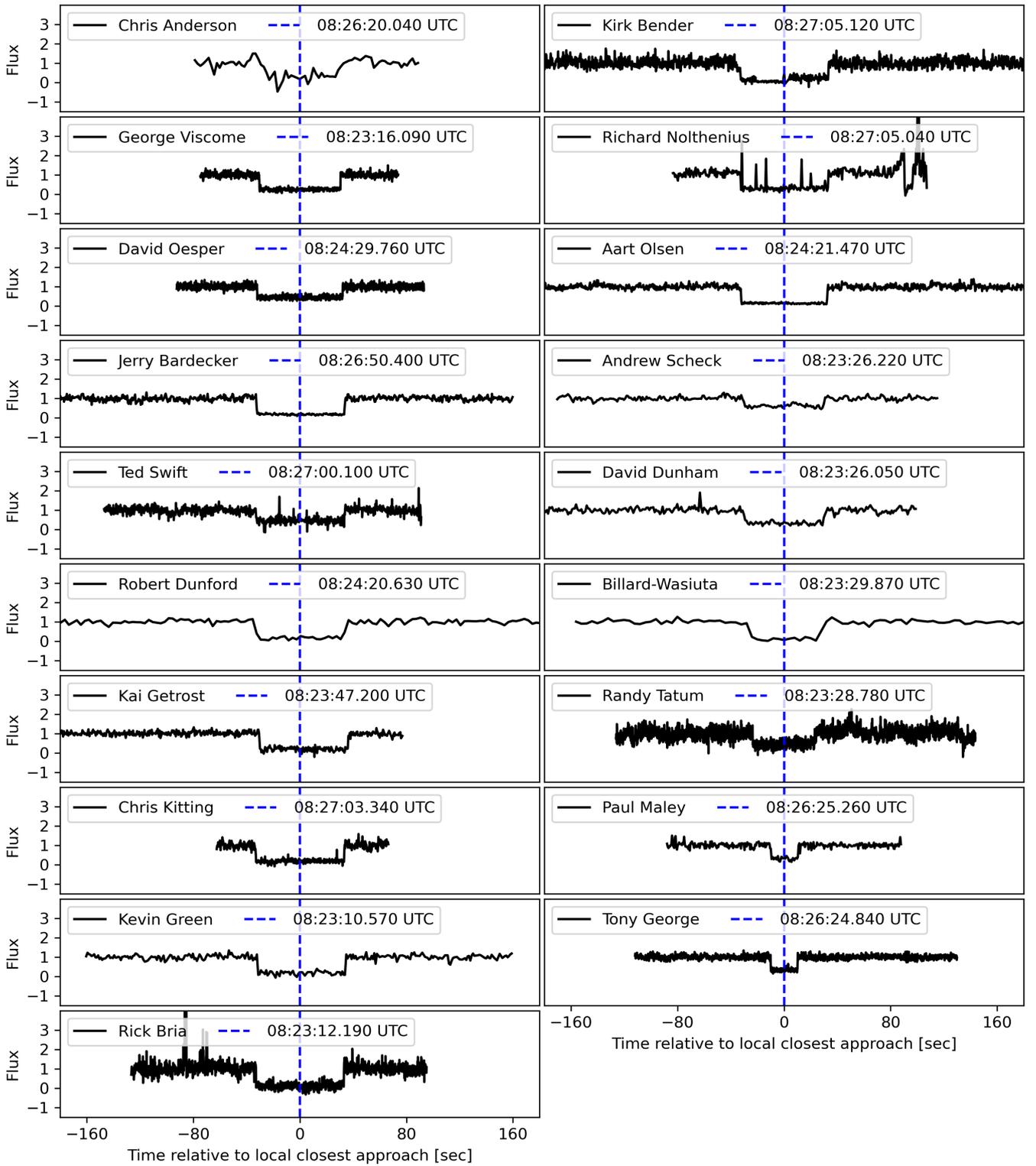

**Figure 3.** All 19 positive light curves from the Umbriel occultation. Flux ratios Umbriel/calibrator were normalized outside the event. Blue lines indicate the displayed local mid-instants of the occultation.

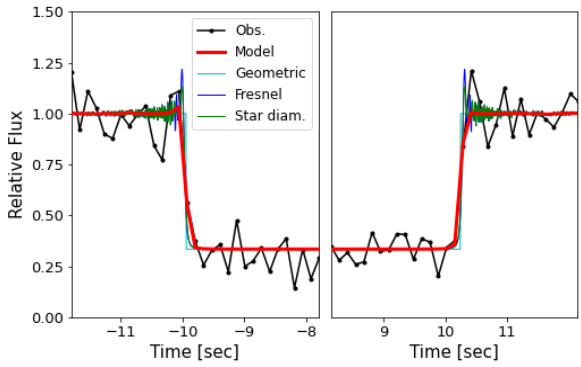

**Figure 4.** Light curve fitting from Tony George. The observed light curve (black dots/lines) was fitted by an iterative $\chi^2$ procedure with synthetic light curve models. The best fitted light curve model is shown in red. The model results from the convolution of a sharp-edge (box geometry) model (light blue) with effects caused by the wavelength-dependent Fresnel diffraction (purple), apparent limb-darkened star diameter (green), and local normal star velocity coupled with the finite exposure and dead (readout) times. The obtained ingress/egress instants and errors come from the geometric sharp-edged occultation model resulting from the fitting. The normalized relative flux is plotted as a function of the seconds before and after the local closest approach (2020-09-21 08:26:24.699 UTC).

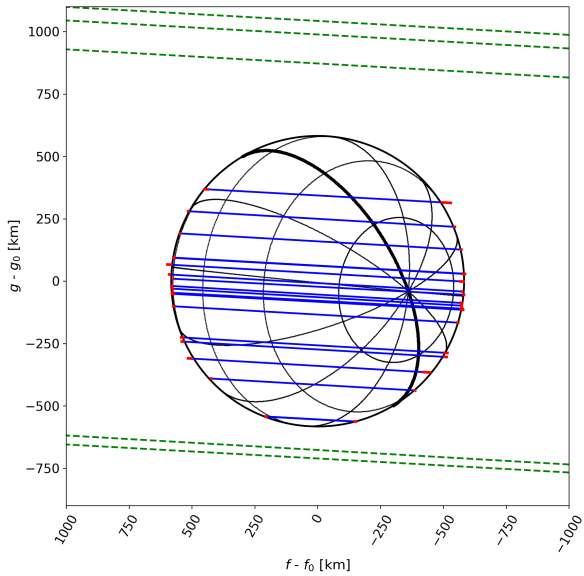

**Figure 5.** 19 positive (blue) chords fitted to a circular limb, and 5 negative (green). The error bar (1 $\sigma$) is represented at the ends of the chords (red). The diameter obtained is 1164.8 km. The apparent limb fit with an ellipse is visually indistinguishable. Up is north and left is east in the $(f,g)$ plane of the sky, with $(f_0, g_0)$ being the ephemeris offsets. The highlighted meridian is in the plane containing Umbriel's poles and Uranus, and indicates where a putative Umbriel ellipsoid would be more oblate, since the satellite is in "tidal-locking" with the planet.

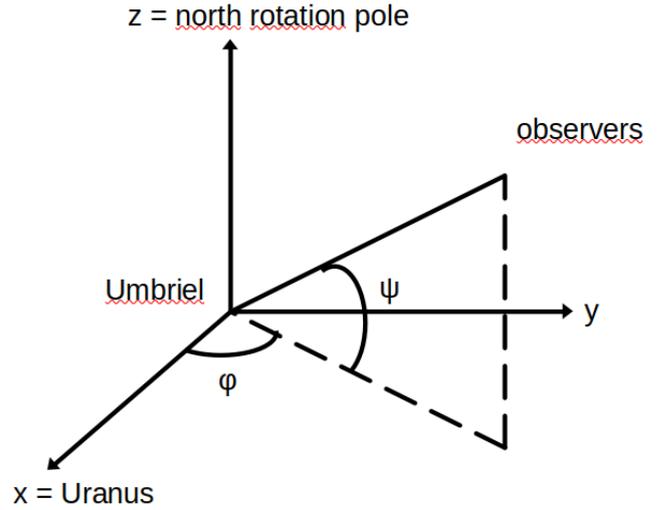

**Figure 6.** Right-handed Umbriel-centered coordinate frame with $x$ and $z$ axes pointing respectively at Uranus and the north rotation pole. Umbriel's equatorial plane $xy$ makes an angle $\psi$ with the direction to the occultation observers, which projected in this plane gives a longitude angle $\varphi$ with respect to the $x$ axis. The angle $\varphi$ is the difference between the sub-observer and sub-Uranus longitude points and $\psi$ is the sub-observer latitude, given in Table 1 at occultation epoch.

### 6.2 Density, surface gravity, escape velocity

Using the fitted radius and error obtained in Section 6.1, we get an spherical volume of $V$ = 827 ± 3 x $10^6$ km$^3$. For a mass of 1.275 ± 0.028 × $10^{21}$ kg (Jacobson 2014, Table 1), we obtain a density $\rho$ = 1.54 ± 0.04 g cm$^{-3}$ for Umbriel. The surface gravity is 0.251 ± 0.006 m s$^{-2}$ and the escape velocity 0.541 ± 0.006 km s$^{-1}$.

### 6.3 Albedo

The geometric albedo $p_V$ is computed by equation 2, where $au$ = 149 597 870.7 km is the Astronomical Unit, $R$ is the body radius in km, and $H$ and $H_\odot$ are the visual absolute magnitudes of the body and the Sun ($H_\odot$ = -26.74), respectively. Karkoschka (2001) gives errors and minimum/maximum values of $H_{max}$ = 1.75 and $H_{min}$ = 1.77 for Umbriel; we adopt here the average value $H$ = 1.76 ± 0.04. Taking the fitted radius and error obtained in Section 6.1, we get a geometric albedo of $p_V$ = 0.26 ± 0.01 for Umbriel.

$$p = \left(\frac{au}{R}\right)^2 10^{0.4(H_\odot - H)} \qquad (2)$$

### 6.4 Limits of an atmosphere on Umbriel

Following Wideman et al.'s detailed discussion (Widemann et al. 2009, Section 6), we consider a $CO_2$ isothermal atmosphere with $T$ = 70 K for Umbriel (Grundy et al. 2006). There are two possible ways to detect or estimate an upper limit for this atmosphere.

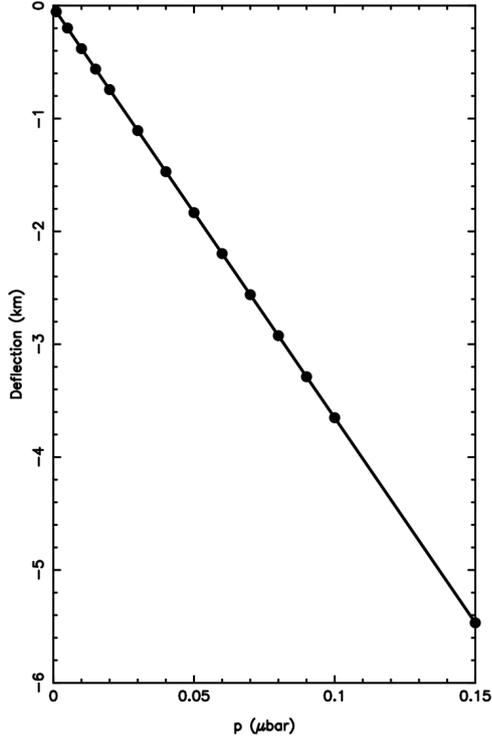

**Figure 7.** Deflection of rays grazing Umbriel's surface as a function of the surface pressure of an isothermal $CO_2$ atmosphere at $T = 70$ K. The $1\sigma$ upper limit of $|R_{occ} - R_{V2}| = 5.2$ km is achieved at pressure $p_{CO2} = 150$ nbar.

### 6.4.1 Atmosphere constraint from radius measurements

Umbriel's radius is known from images taken during the Voyager 2 flyby in 1986, with a value $R_{V2} = 584.7 \pm 2.8$ km (Thomas 1988). It is then possible to compare $R_{V2}$ with the radius $R_{occ}$ obtained from the occultation. An atmosphere will slightly shrink the shadow radius due to the ray bending caused by refraction, so that $R_{occ} - R_{V2}$, corresponding to that deflection, is expected to be negative.

The present work yields $R_{occ} = 582.4 \pm 0.8$ km, so that $R_{occ} - R_{V2} = -2.3 \pm 2.9$ km. Thus, no significant shrinking is detected, only a $1\sigma$ upper limit of $|R_{occ} - R_{V2}| < 5.2$ km can be given.

Fig. 7 shows the deflection $R_{occ} - R_{V2}$ as a function of the surface pressure $p_{\mu bar}$ of a $CO_2$ isothermal atmosphere with $T = 70$ K. This plot shows that a $1\sigma$ upper limit of $p_{CO2} = 150$ nbar is obtained by this method.

### 6.4.2 Atmosphere constraint from light curves

An atmosphere causes a drop of signal for several tens of kilometers above the surface. Synthetic light curves – obtained as in Widemann et al. (2009) – can then test the possibility of an atmosphere.

We have generated such synthetic light curves for surface pressures ranging from 1 nbar to 200 nbar. For each model, we have calculated the $\chi^2$ value of the model by equation 3, where $N$ is the number of data points, $\Phi_i$ is the observed flux of the $i^{th}$ data point, $\Phi_{i,synth}$ is the corresponding synthetic flux obtained from the model, and $\sigma$ is the standard deviation of the observed flux.

$$\chi^2 = \sum_{1}^{N} \left( \frac{\Phi_i - \Phi_{i,synth}}{\sigma} \right)^2 \qquad (3)$$

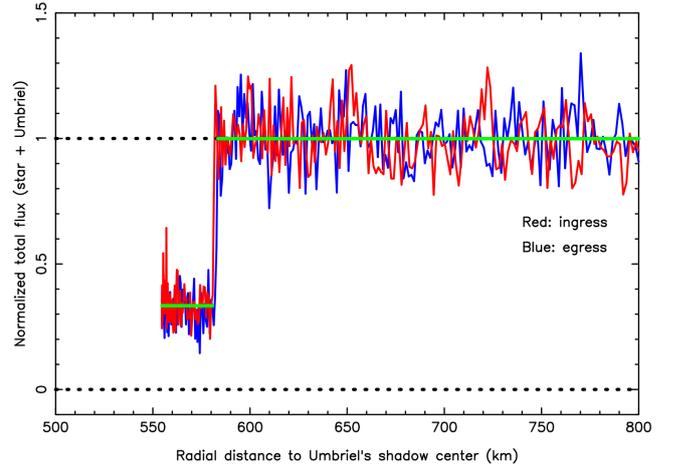

**Figure 8.** Fittings (green) of the baseline $\phi_0 = 0.3335$ and full unocculted flux ratio $\phi_1 = 0.9996$ of the Tony George light curve. Each light curve point is expressed by its radial distance (km) to Umbriel's shadow center. Red/blue refer to ingress/egress points. A rms deviation $\sigma = 0.0947$ is obtained outside the occultation. The radius range $r$ of points used for computing $\phi_0$ was $0 < r < 581$ km, while for $\phi_1$ we used 583 km $< r < \sim 2000$ km. Values $\phi_0$ and $\phi_1$ are used to generate synthetic light curves, and $\sigma$ used in the atmosphere $\chi^2$ fittings.

The best observed light curve in terms of signal-to-noise ratio and resolution was obtained from Tony George (Tables 2 and 3). The fittings of the baseline $\phi_0 = 0.3335$ and unocculted flux $\phi_1 = 0.9996$ of this light curve are shown in Fig. 8. The light curve rms deviation was $\sigma = 0.0947$. The $\phi_0$ and $\phi_1$ values were used to generate the synthetic light curve fluxes $\Phi_{synth}$ of the putative atmosphere, and $\sigma$ was used for the calculation of the $\chi^2$ values of the atmospheric fittings.

Fig. 9 shows $\chi^2$ vs. various surface pressures. The monotonic variation of $\chi^2$ indicates that no atmosphere is detected. Taking the criterion $\chi^2 + 1$ as a $1\sigma$ upper limit, we obtain $p_{CO2} < 13$ nbar ($1\sigma$). Using the $\chi^2 + 9$ criterion, we obtain a $3\sigma$ upper limit for the surface pressure of $p_{CO2} < 72$ nbar ($3\sigma$). Fig. 10 shows how a $p_{CO2} = 72$ nbar atmosphere fits the observed Tony George light curve.

The 13 nbar-limit is much larger than the equilibrium vapour pressure expected for a $CO_2$ atmosphere at 70 K, which amounts to $p_{sat}(CO_2) = 0.16 \times 10^{-3}$ nbar, see Fray & Schmitt (2009). This is fully consistent with the absence of detection of a global $CO_2$ atmosphere in our data set. However, as detailed by Widemann et al. (2009), this does not preclude a local atmosphere and significant sublimation-condensation cycles and seasonal redistribution, as shown by Grundy et al. (2006). In particular, due to the very steep dependence of $p_{sat}(CO_2)$ upon temperature, a value of 13 nbar is reached for a surface temperature of $T \approx 92$ K. Although no such local temperatures have been reported for Umbriel, we note that subsolar temperatures as high as 86 K have been mentioned for other Uranian satellites (Hanel et al. 1986). Note also that Umbriel has spots such as the bright ring on the floor of 130 km crater Wunda, similar to the ring on the floor of Pluto's Elliot crater, which may be another example of volatile accumulation in topographic lows (Schenk & Moore 2020).

Thus, although the absence of a global Umbriel's $CO_2$ atmosphere is expected from the beginning, our data serve to evaluate a local

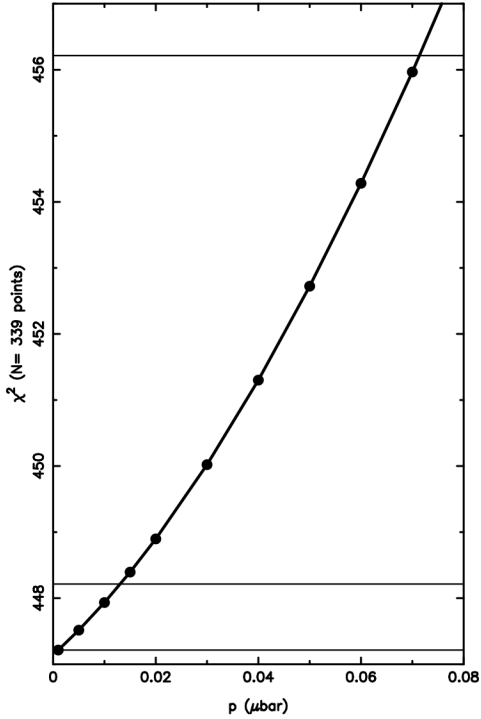

**Figure 9.** The $\chi^2$ value obtained from synthetic light curves using an isothermal $CO_2$ atmospheres with various surface pressures and $N$ = 339 data points. The lowest horizontal solid line is the minimum value $\chi^2_{min}$ of $\chi^2$, the middle solid line marks the $1\sigma$ limit corresponding to $\chi^2_{min}$ + 1 and $p_{CO2}$ = 13 nbar, while the top horizontal line marks the $3\sigma$ limit $\chi^2_{min}$ + 9 with $p_{CO2}$ = 72 nbar.

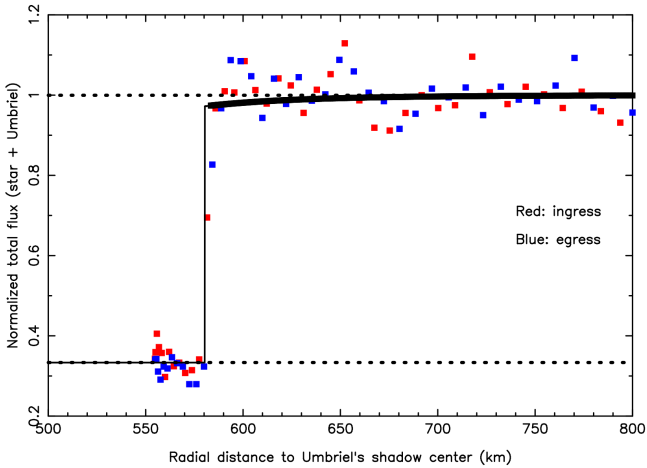

**Figure 10.** Light curve of Tony George binned over 6 data points, yielding a radial separation of about 4.5 km between two consecutive data points (radius in km). The dotted lines are the fitted baseline and full unocculted fluxes. The solid curve is a synthetic model showing the effect of an isothermal $CO_2$ atmosphere with a surface pressure of 72 nbar, corresponding to the $3\sigma$ limit of detection for such an atmosphere (see Fig. 9).

nbar-level atmosphere that might exist over a warmer spot along the sampled Umbriel's limb.

### 6.5 Milliarcsecond precision level position for Umbriel

From the sub-mas precision Gaia DR3 position (Table 1) and the high precision ($f_c$,$g_c$) ephemeris offsets obtained from the circular limb fittings (Table 4), we derived an accurate/precise geocentric position for Umbriel at the occultation epoch September 21st, 2020, 08 h 24 m 36.000s UTC (equation 4). By virtue of this technique, ephemeris errors actually do not enter the occultation-based position error budget. Ephemeris offsets are given in the sense "Occultation minus DE435/URA111", with errors coming from the fitted ($f_c$,$g_c$).

$$\alpha = 02^h 30^m 28.^s 84556 \pm 0.1 mas$$
$$\delta = 14^o 19' 36''.5836 \pm 0.2 mas$$
$$\Delta\alpha \, cos\delta = -1.7 \pm 0.1 mas$$
$$\Delta\delta = -12.3 \pm 0.2 mas \qquad (4)$$

## 7 DISCUSSION AND CONCLUSIONS

Table 5 summarizes Umbriel's physical characteristics obtained from the September 21st, 2020 stellar occultation observations analysed in the previous sections. For comparison, results from Voyager II's flyby on January 1986 are also given.

Umbriel is expected to be an ellipsoid with roughly equal semi-minor axes, and semi-major axis pointed at Uranus. However, as pointed out in Section 6.1, we could not derive a robust elliptical limb solution despite the kilometric precision of the results. This is probably caused by the high positive/negative amplitudes at the 5 km level of Umbriel's limb, already measured by Voyager II and discussed below. Therefore, we adopt a spherical limb as the main solution and interpret the elliptical results as upper limits for the putative Umbriel ellipsoid.

The smaller and more precise occultation radius gives Umbriel a bit higher, more precise density of $\rho$ = 1.54 ± 0.04 g cm$^{-3}$ than the previous determination of $\rho$ = 1.39 ± 0.16 g cm$^{-3}$ by Jacobson (1992b). The geometric albedo is compatible with the 0.258 value furnished in Table VII of Karkoschka (2001).

Analysing Voyager II images acquired during its visit to the system in January 1986, Thomas (1988) could not derive axis sizes for Umbriel, Titania or Oberon. The aspect angle of the illuminating Sun with respect to the visible south pole of the system was roughly $8^o$, and a mean radius of 584.7 ± 2.8 km was obtained for Umbriel. At the time of the occultation, the aspect angle of the visible north pole to the occultation observers was about $38^o$ (Table 1), and the fitted circular limb gives a somewhat smaller and more precise radius of 582.4 ± 0.8 km.

Upper limits derived from the occultation for the true semi-axes of a putative Umbriel ellipsoid result in an equivalent radius of 583.6 ± 2.2 km. The difference from Voyager II radius – obtained when the full satellite's equator limb was accessible from the south – is 1.1 ± 3.6 km, with a good agreement within the errors. Using the more precise spherical solution results in an "occultation minus Voyager II" radius difference of -2.3 ± 2.9 km – yet not a significant difference, with a corresponding $1\sigma$ upper limit of $|R_{occ} - R_{V2}|$ < 5.2 km.

If the radius difference is due to the shrinking effect of a tenuous, putative isothermal $CO_2$ atmosphere at $T$ = 70 K, a $1\sigma$ upper limit of 5.2 km corresponds to a surface pressure of less than 150 nbar. A much more sensitive method for detecting an atmosphere comes

**Table 5.** Umbriel physical characteristics and astrometry from the September 21th, 2020 stellar occultation. Voyager II results are also listed for comparison.

| Umbriel physical characteristics from occultation | |
|---|---|
| Radius | $R$ = 582.4 ± 0.8 km |
| Limb parameter | $\sigma_{model}$ = 4.2 km |
| Density (sphere) | $\rho$ = 1.54 ± 0.04 g cm$^{-3}$ |
| Surface gravity | 0.251 ± 0.006 m s$^{-2}$ |
| Escape velocity | 0.541 ± 0.006 km s$^{-1}$ |
| Geometric albedo (sphere) | $p_V$ = 0.26 ± 0.01 |
| Aspect angle | $\zeta$ = 38$^o$ 16' 24''.82 (north pole) |
| Isothermal CO$_2$ atmosphere | surface pressure = 13 – 72 nbar |
| at $T$ = 70 K | (1–3$\sigma$ upper limits) |
| Upper limits on putative Umbriel ellipsoid from occultation | |
| True semi-major axis | $a$ = 584.9 ± 3.8 km |
| True semi-minor axes | $b = c$ = 582.3 ± 0.6 km |
| True oblateness | $\epsilon$ = 0.004 ± 0.008 |
| True equivalent radius [1] | $R_{eq}$ = 583.6 ± 2.2 km |
| Astrometry occultation data | |
| Ephemeris offsets [2] | $\Delta\alpha \cos\delta$ = -1.7 ± 0.1 mas |
| (mas) | $\Delta\delta$ = -12.3 ± 0.2 mas |
| Ephemeris offsets [2] | $\Delta\alpha \cos\delta$ = -23.7 ± 0.8 km |
| (km) | $\Delta\delta$ = -169.4 ± 1.9 km |
| Geocentric ICRS | $\alpha$ = 02$^h$ 30$^m$ 28.$^s$84556 ± 0.1 mas |
| position at epoch | $\delta$ = 14$^o$ 19' 36''.5836 ± 0.2 mas |
| Reference occultation epoch | 2020-09-21 08:24:36.000 UTC |
| Voyager II 1986 observations | |
| Radius (Voyager II, 1986) | $R$ = 584.7 ± 2.8 km (Thomas 1988) |
| Limb topography (Voyager II, 1986) | ± 5.0 km (Fig. 3 in Thomas 1988) |
| Aspect angle with the Sun [3] | $\zeta$ = 8$^o$ (south pole) |

*Notes.* (1) For an ellipsoid with true semi-major $a$ and true oblateness $\epsilon$ with semi-minor axes $b = c$, the true equivalent radius $R_{eq}$ corresponds to the radius of a circle with the same area of the ellipse limb for an edge-on sight of the ellipsoid, i.e. $R_{eq} = a\sqrt{1-\epsilon}$. (2) Ephemeris offsets in the sense "Occultation minus DE435/URA111". (3) More than the spacecraft one, the aspect angle with the Sun indicates the illuminated surface of Umbriel which could be actually analysed in Voyager II images.

from analysing high signal-to-noise ratio light curves (Section 6.4.2). Using the Tony George curve, we derive more precise estimates for the surface pressure of an isothermal CO$_2$ atmosphere at $T$ = 70 K: 13 – 72 nbar for 1–3$\sigma$ upper limits.

The equilibrium vapour pressure expected for a CO$_2$ atmosphere at 70 K is much smaller, $p_{sat}$(CO$_2$) = 0.16×10$^{-3}$ nbar (Fray & Schmitt 2009). However, our data is sensitive enough for the detection of a local nbar-level atmosphere. The 1$\sigma$ upper limit for the surface pressure obtained here is similar to the limit obtained for Titania (also 13 nbar, Widemann et al. 2009). If vapour pressure is attained, then p$_{CO_2}$ = 13 nbar would require a surface temperature of about 92 K, a bit warmer than the ∼ 60-80 K range estimated for the mean surface temperature of Umbriel by Grundy et al. (2006). Thus, in this case, a 13 nbar CO$_2$ atmosphere could be maintained by sublimation of some hotter spots on the surface of the satellite. Other gases such as N$_2$, CH$_4$ and CO are not considered here. They are very volatile, and would reach unrealistic pressures well above 1 mbar. This would be very conspicuous in the analysed light curve, and more importantly, it would lead to a massive thermal evaporation and a rapid loss of such atmospheres (Widemann et al. 2009). Conversely, water ice is too involatile to produce any atmosphere at the temperatures considered here.

A limb parameter ($\sigma_{model}$) of 4.2 km was obtained for the circular limb fitting (4.3 km for the elliptical one) for an aspect angle of 38$^o$ with the north pole. It is in remarkable agreement with the ± 5 km limb variations reported by Thomas (1988) from the analysis of the best Voyager II images, taken at a distinct aspect angle – near pole-on from the south. Improved image processing including photoclinometry techniques have recently been applied to enhance the resolution of Voyager II images (Schenk & Moore 2020). Numerous individual relief features (mostly craters) spread all over the southern hemisphere could finally have their diameters, heights and depths measured, confirming the strong limb variation with very consistent values to our fitted 4.2 km limb parameter. All this indicates that both southern and northern Umbriel hemispheres share a most irregular surface with the same depth/height 4 km scale – by far the largest topographic variation among the main Uranian moons, comparable to Miranda.

## ACKNOWLEDGEMENTS

This research made use of SORA, a Python package for stellar occultation reduction and analysis, developed with the support of ERC Lucky Star and LIneA/Brazil. This study was financed in part by the National Institute of Science and Technology of the e-Universe project (INCT do e-Universo, CNPq grant 465376/2014-2). The following authors acknowledge the respective CNPq grants: M.A. 427700/2018-3, 310683/2017-3, 473002/2013-2; B.E.M. 150612/2020-6; F.B.R. 314772/2020-0; G. M. 128580/2020-8; J.I.B.C. 305917/2019-6, 306691/2022-1; R.V-M 307368/2021-1. J.I.B.C. also acknowledges FAPERJ (grant 201.681/2019). G. M. also thanks CAPES for scholarship 88887.705245/2022-00. ARGJr acknowledges FAPESP grant 2018/11239-8. G.B-R. acknowledges CAPES - FAPERJ/PAPDRJ grant E26/203.173/2016 and the scholarship granted in the scope of the Program CAPES-PrInt, process number 88887.310463/2018-00, Mobility number 88887.571156/2020-00. This work has made partial use of data from the European Space Agency (ESA) mission Gaia (https://www.cosmos.esa.int/gaia), processed by the Gaia Data Processing and Analysis Consortium (DPAC, https://www.cosmos.esa.int/web/gaia/dpac/consortium). This research has made partial use of the VizieR catalogue access tool, CDS, Strasbourg, France (https://cds.u-strasbg.fr/vizier-org/licences_vizier.html). The original description of the VizieR service was published in Ochsenbein et al. (2000).

## DATA AVAILABILITY

The data that support the results and plots in this paper and other findings of this study are available from the corresponding author upon reasonable request.

This paper has been typeset from a TeX/LaTeX file prepared by the author.